\documentclass[]{aa}

\usepackage[normalem]{ulem}
\usepackage{graphicx}
\usepackage{txfonts}
\usepackage{natbib}
\bibpunct{(}{)}{;}{a}{}{,}
\usepackage{scrextend}		
\usepackage{tikz}
\usepackage{subcaption}
\usepackage{multirow}
\usepackage{hyperref}
\usepackage{cleveref}
\usepackage{amsmath}  
\usepackage{xfrac}   
\usepackage{amssymb}	
\usepackage{graphicx}
\usepackage{amsmath}  
\usepackage{xfrac}   
\usepackage{amssymb}	
\usepackage{dsfont}
\usepackage{txfonts}
\usepackage{mwe,tikz}\usepackage[percent]{overpic}

\newcommand{\mdotin}{$\dot{m}_{in}$\xspace}

\newcommand{\rj}{$r_J$\xspace}
\newcommand{\pair}{$(r_J, \dot{m}_{in})$\xspace}
\newcommand{\fig}{Fig.\xspace}

\newcommand{\eg}{e.g.\xspace}

\newcommand{\gx}{GX~339-4\xspace}
\newcommand{\xspec}{\textsc{xspec}\xspace}
\newcommand{\HT}{hard tail\xspace}

\begin{document}

  \title{A unified accretion-ejection paradigm for black hole X-ray binaries}

  \subtitle{IV. Replication of the 2010--2011 activity cycle of \gx }

  \author{G. Marcel\inst{1} \fnmsep \inst{2}
          \and
          J. Ferreira\inst{1} 
           \and
          M. Clavel\inst{1} 
          \and 
          P-O. Petrucci\inst{1}
          \and
          J. Malzac\inst{3}
          \and
          S. Corbel\inst{4} \fnmsep \inst{5}
          \and
          J. Rodriguez\inst{4}
          \and
          R. Belmont\inst{3} \fnmsep \inst{4}
		 \and
          M. Coriat\inst{3}
          \and
          G. Henri\inst{1}
          \and
          F. Cangemi\inst{4}
   		}

   \institute{Univ. Grenoble Alpes, CNRS, IPAG, 38000 Grenoble, France \\
              \email{gregoiremarcel26@gmail.com or gregoire.marcel@villanova.edu}
              \and 
              Villanova University, Department of Physics, Villanova, PA 19085, USA
              \and
              IRAP, Universite de Toulouse, CNRS, UPS, CNES, Toulouse, France
              \and
              AIM, CEA, CNRS, Université Paris-Saclay, Université Paris Diderot, Sorbonne Paris Cité, F-91191 Gif-sur-Yvette, France
              \and 
              Station de Radioastronomie de Nançay, Observatoire de Paris, PSL Research University, CNRS, Univ. Orléans, 18330 Nançay, France
             }

   \date{Received 15 January 2019; accepted 10 May 2019}

   \abstract
   {Transients X-ray binaries (XrB) exhibit very different spectral shapes during their evolution. In luminosity-color diagrams, their behavior in X-rays forms q-shaped cycles that remain unexplained. We proposed a framework where the innermost regions of the accretion disk evolve as a response to variations imposed in the outer regions. These variations lead not only to modifications of the inner disk accretion rate $\dot m_{in}$ but also to the evolution of the transition radius $r_J$ between two disk regions. The outermost region is a standard accretion disk (SAD), whereas the innermost region is a jet-emitting disk (JED) where all the disk angular momentum is carried away vertically by two self-confined jets.}
   {In the previous papers of this series, it has been shown that such a JED--SAD disk configuration could reproduce the typical spectral (radio and X-rays) properties of the five canonical XrB states. The aim of this paper is now to replicate all X-ray spectra and radio emission observed during the 2010--2011 outburst of the archetypal object \gx .}
   {We use the two--temperature plasma code presented in two previous papers (II, III) and design an automatic ad hoc fitting procedure that gives, for any given date, the required disk parameters $(\dot m_{in},r_J)$ that best fit the observed X-ray spectrum. We use X-ray data in the 3--40~keV (\textit{RXTE}/PCA) spread over 438 days of the outburst, together with 35 radio observations at 9~GHz (ATCA) dispersed within the same cycle.}
   {We obtain the time distributions of $\dot m_{in}(t)$  and $r_J(t)$ that uniquely reproduce the X-ray luminosity and the spectral shape of the whole cycle. Using the classical self-absorbed jet synchrotron emission model, the JED--SAD configuration reproduces also very satisfactorily the radio properties, in particular the switch-off and -on events and the radio-X-ray correlation. Although the model is simplistic and some parts of the evolution still need to be refined, this is, to our knowledge, the first time that an outburst cycle is reproduced with such a high level of detail.}
   {Within the JED--SAD framework, radio and X-rays are so intimately linked that radio emission can be used to constrain the underlying disk configuration, in particular during faint hard states. If this result is confirmed using other outbursts from \gx or other X-ray binaries, then radio could be indeed used as another means to indirectly probe disk physics.}

   \keywords{black hole physics --
                accretion, accretion discs --
                magnetohydrodynamics (MHD) -- 
                ISM: jets and outflows --
                X-rays: binaries
               }

   \maketitle
%

\section{Introduction}

The time and spectral behaviors of transient X-ray binaries are important challenges for the comprehension of the accretion-ejection phenomena. These binary systems can remain in quiescence for years before suddenly going into outburst, usually for several months. During a typical outburst cycle, the mass accretion rate onto the central compact object undergoes a sudden rise, leading to an increase in X-ray luminosity by several orders of magnitude, before decaying back to its initial value. These two phases are referred to as the rising and decaying phases. The X-ray spectrum is also seen to vary significantly during these events, displaying two very different spectral shapes. It is either dominated by a hard power-law component above 10~keV (defined as the hard state), or dominated by a soft black-body component of a few keV (soft state). During the rising phase, all objects display hard state spectra, until at some point they transition to a soft state. When transitioning, a significant decrease in luminosity is undergone before coming back to the hard state. There is therefore a striking hysteresis behavior: XrB transients show two very different physical states, and the two transitions from one state to another happen at different luminosities. This provides the archetypal 'q' shaped curve of X-ray binaries in the so-called hardness-intensity diagram. An evolutionary track for which no satisfactory explanation for state transitions has been provided yet \citep{Remillard06}. For recent reviews and surveys, we refer the reader to, for example, \citet{Dunn10} or \citet{Tetarenko16}.

In addition to specific accretion cycles, X-ray binaries also show specific radio properties (jets) correlated to the X-ray behavior \citep{Corbel03,Gallo03}. This puzzling fact was already noted in early studies \citep[see, e.g.,][]{HW71,Tananbaum72,Bradt75}. Indeed, persistent self-collimated jets, as probed by a flat-spectrum radio emission \citep{BK79}, are detected during hard states, whereas no radio emission is seen during soft states. This defines thereby an imaginary line where jets are switched-off: the so-called "jetline" \citep{Fender09}, also marking the moment where discrete ejections of plasma bubbles are observed \citep[see for example][]{1998Natur.392..673M, Rodriguez08}, and after which sources are in radio quiet states. Nowadays, it is widely accepted that spectral changes are due to modifications in the inner accretion flow structure \citep[see, e.g.,][and references therein]{Done07}, and that detection/non-detection of radio emission results from the presence/absence of compact jets (\citeauthor{Corbel04} \citeyear{Corbel04}, \citeauthor{Fender04} \citeyear{Fender04}, see however \citeauthor{Drappeau17} \citeyear{Drappeau17} for an alternative view).

A global scenario was first\footnote{We also refer an interested reader to \citet{Lasota96} for an ealier, yet different, global view.} proposed by \citet{Esin97}, based on the interplay between an outer standard accretion disk \citep[SAD hereafter,][]{SS73} and an inner advection-dominated flow \citep{Ichimaru77, Rees82, Narayan94}. While the presence of a SAD in the outer disk regions is globally accepted \citep{Done07}, the existence and the physical properties of the inner flow remain highly debated for X-ray binaries \citep[for a review, see][]{Yuan14}. This scenario does not however address the jet formation and quenching, leaving an important observational diagnostic unexplained. 

A framework addressing the full accretion-ejection phenomena has been proposed and progressively elaborated in a series of papers. \citet{Ferreira06}, hereafter paper~I, proposed that the inner disk regions would be threaded by a large-scale vertical magnetic field. Such a $B_z$ field is assumed to build up mostly from accumulation from the outer disk regions, as seen in very recent numerical simulations \citep[see, e.g.][]{2018arXiv180904608L}. As a consequence, its radial distribution and time evolution are expected to vary according to the (yet unknown) interplay between advection by the accreting plasma and the turbulent disk diffusion. The local field strength is then measured at the disk mid-plane by the magnetization $\mu(r) =B_z^2/P$, where $P$ is the total (gas plus radiation) pressure. The main working assumption of this framework is that the magnetization increases inwardly so that it reaches a value allowing a jet-emitting disk (hereafter JED) to establish. 

The properties of JEDs have been extensively studied, mostly analytically \citep[]{Ferreira93a,Ferreira95,Ferreira97,Casse00b} but also numerically \citep[]{CasseKeppens02,Zanni07,Murphy10,Tzeferacos13}. In these solutions, all the disk angular momentum and a sizable fraction of the released accretion power are carried away by two magnetically-driven jets \citep{BP82}. These jets produce an important torque on the underlying disk, allowing accretion to proceed up to supersonic speeds. This characteristic and quite remarkable property stems from the fact that JEDs require near equipartition $B_z$ field, namely $\mu$ lying roughly between 0.1 and 0.8. As a consequence, a JED becomes sparser than a SAD fed with the same mass accretion rate.

\citet{Marcel18a}, hereafter paper~II, developed a two-temperature plasma code that computes the local disk thermal equilibria, taking into account the advection of energy in an iterative way. The code addresses optically thin/thick transitions, both radiation and gas supported regimes, and computes in a consistent way the emitted global spectrum from a steady-state disk. The optically thin emission is obtained using the \textsc{Belm} code \citep{Belmont08,Belmont09}, a code that provides accurate spectra for bremsstrahlung and synchrotron emission processes, as well as for their local Comptonization. It turns out that JEDs, because of their low density even at high accretion rates, naturally account for luminous hard states with luminosities up to $30\%$ the Eddington luminosity \citep{Eddington}, a level hardly achieved in any other accretion mode \citep{Yuan14}. 

However, a disk configuration under the sole JED accretion mode cannot explain cycles such as those exhibited by \gx . Not only does the system need to emit an X-ray spectrum soft enough when transitioning to the soft state, but jets need also to be fully quenched. To do so, we assume the existence of a transition at some radius $r_J$, from an inner JED to an outer SAD, as proposed in paper~I. This results in an hybrid disk configuration raising several additional difficulties in the treatment of the energy equation. One of them is the non local cooling of the inner (usually hot) JED by soft photons emitted by the external SAD, another is the advection of colder material into the JED. Both effects have been dealt with in \citet{Marcel18b}, hereafter paper~III. They explored the full parameter space in disk accretion rate and transition radius, and showed that the whole domain in X-ray luminosities and hardness ratios covered by standard XrB cycles is well reproduced by such hybrid disk configurations. Along with these X-ray signatures, JED--SAD configurations also naturally account for the radio emission whenever it is observed. As an illustration, five canonical spectral states typically observed along a cycle were successfully reproduced and displayed.

In this paper, we make the next step. We show that a smooth evolution of both the inner disk accretion rate $\dot m_{in}(t)$ and transition radius $r_J(t)$ can simultaneously reproduce the X-ray spectral states and the radio emission of a typical XrB, \gx , during one of its outbursts. 
In section~\ref{sec:GX}, we present the observational data used in this article, in X-rays and at radio wavelengths. Then, in section~\ref{sec:reprod1011}, we present the fitting procedure implemented to derive the best pair of parameters \pair allowing to reproduce the evolution of the X-ray (3--40~keV) spectral shape. Although we focused only on X-rays, the model predicts a radio light curve that is qualitatively consistent with what is observed. It turns out that the phases within the cycle with the largest discrepancies are also those where the constraints imposed by X-rays are the loosest. We thus included the radio (9~GHz) constraints within the fitting procedure, leading to a satisfactory quantitative replication of both the X-ray and radio emission along the whole cycle (Sect.~\ref{sec:Xray+R}). Section~5 concludes by summarizing our results.

\section{Spectral and radio evolution of GX~339-4} 
\label{sec:GX}

\subsection{Data selection and source properties}

In order to investigate the capability of our theoretical model to reproduce the outbursts of X-ray binaries, we need a large number of observations tracing the spectral evolution of an X-ray binary through a given outburst. Among all X-ray observations available, we therefore selected the \textit{RXTE} archival data, which currently provide the best coherent coverage of such outbursts. In the past century, \gx was one of the first X-ray sources discovered. 
Since then, \gx has been widely studied and has shown to be one of the most productive X-ray binaries, undergoing an outburst once every two years on average \citep[see][Table 14 for a complete review]{Tetarenko16}. For this historic reason and the huge amount of data available, we will focus our study on this notorious object. The 2010--2011 outburst of \gx was then chosen because it has the best simultaneous radio coverage.

Among the parameters of \gx , the distance to the source seems to be the best constrained. Different studies have estimated with accurate precision that $d \simeq 8 \pm 1$~kpc \citep{Hynes04, Zdz04, Parker16}. The spin of the central black hole of \gx has also been constrained using different spectral features, and the most recent studies seem to agree with values such as $a \simeq 0.93$--$0.95$ \citep[][but see \citeauthor{Ludlam15} \citeyear{Ludlam15} for different and higher spin estimates $a>0.97$.]{Reis08, Miller08, Garcia15}. The spin of the black hole is not a direct parameter within our model, but the inner radius of the disk is. We assume that the disk extends down to the inner-most stable circular orbit (ISCO) of the black hole. Thus, in agreement with the estimation of the spin, we choose $r_{in} = R_{in} / R_g = 2$, i.e. $a=0.94$, where $R_g = GM/c^2$ is the gravitational radius, $G$ the gravitational constant, $c$ the speed of light, and $M$ the black hole mass. An important parameter of our model is obviously the black hole mass $m = M/M_{\odot}$, where $M_{\odot}$ is the mass of the Sun. Multiple studies have been performed to constrain $m$ and have led to different estimations \citep[see for example,][]{Hynes03,Munoz08,Parker16,Heida17}. For simplification, and since there is no consensus on \gx mass, we chose it to be a rather central value, the same as the former mass function $m = 5.8$. In any case, the self-similar modelling implies that our results are insensitive to the uncertainties on black hole mass. The inclination of the system is neglected for now (see papers II and III).

In this work, cylindrical distances $R$ will be expressed with respect to the gravitational radius $r=R/R_g$, the mass with respect to the solar mass $m=M/M_{\odot}$, luminosities will be normalized to the Eddington luminosity $L_{Edd}$, and the disk accretion rate with $\dot{m} = \dot{M} / \dot{M}_{Edd} = \dot{M} c^2 / L_{Edd}$. In practice, we use only the accretion rate at the innermost disk radius $\dot{m}_{in}= \dot{m}(r_{in})$. Note that this definition of $\dot{m}$ does not include any accretion efficiency. 

\subsection{X-ray observations and spectral analysis}
\label{sec:xraydata}
The spectral analysis of the X-ray observations was restricted to the 3--40~keV energy range covered by \textit{RXTE}/PCA \citep[for more details on data reduction and spectral analysis, see][]{Clavel16}. The best fits obtained included an absorbed power-law (hard X-rays) plus a disk (soft X-rays), providing the power-law photon index $\Gamma$ and luminosity $L_{\rm pl}$ as well as the overall luminosity $L_{\rm 3-200}=L_{\rm pl} + L_{\rm disk}$. The typical maximal statistical errors in \citet{Clavel16} fits are few percent in flux on average: $2\%$ in the soft state, and up to $5\%$ in the hard state. In power-law dominated states, an average statistical error of $\Gamma_{err} \simeq 0.04$ in power-law index was found, whereas this error is $\Gamma_{err} \simeq 1.1$ in pure soft states. Those errors are reported in Fig.~\ref{fig:Big} and Fig.~\ref{fig:ProcB_res}. Note that observed luminosities, $L_{\rm 3-200}$, $L_{\rm pl}$, and $L_{\rm disk}$ are computed in the 3--200~keV range using the power-law and disk parameters fitted between 3--40~keV and extrapolated up to 200~keV. 
 
These spectral fits allowed us to derive several important observational quantities for each \textit{RXTE} observation and to follow their evolution along the 2010--2011 outburst, observed from MJD 55208 (January 12, 2010) to MJD 55656 (April 5, 2011). In Fig~\ref{fig:Evolution}, we display the disk fraction luminosity diagram \citep[DFLD, see][]{Kording06} of \gx in the 3--200~keV range. \gx follows the usual q-shaped hysteresis cycle, crossing all\footnote{During the 2010--2011 outburst, \textit{RXTE} monitoring started when \gx was already in the low-hard (LH) state, which explains the lack of observations during the transition from quiescence (Q) to LH.} five canonical states of X-ray binaries, as defined in paper~III: quiescent, low-hard, high-hard, high-soft and low-soft (hereafter Q, LH, HH, HS and LS, respectively). In paper III, we demonstrated that our model was able to replicate the generic properties observed in these five states. The jet-line \citep{Corbel04, Fender04} overlaid on the DFLD indicates the separation between states associated with a steady\footnote{Steady emission related to the interaction of jets with the interstellar medium is not considered here.} radio emission (right) and those with flares or undetectable radio emission (left), see also Fig.~\ref{fig:Radio}. From now on, days are expressed in reference to MJD~55208 $\equiv$ day~0.

In a pioneering work, \citet{Markert73b} and \citet{Markert73a} discovered that \gx had undergone three different spectral states between 1971 and 1973. They named these states upon their 1--6~keV fluxes: \textit{high-state}, \textit{low-state} and \textit{off-state}. They had just discovered the ancestors of the \textit{soft}, \textit{hard}, and \textit{quiescent} states respectively. In their influential paper, \citet{Remillard06} defined another spectral state localized between the low/hard and the high/soft states: the steep power-law state. They also renamed the high/soft state into thermal state, leading to 3 stable states (hard, steep power-law, thermal). Within this new nomenclature, an object spends most of its life in the quiescent state, before rising in the hard state. Then, it transits from the hard to the steep power-law state in the hard-to-soft transition (upper horizontal branch in the DFLD), and the other way around in the soft-to-hard transition (lower branch). Whenever the object is not in one of the 3 stable states defined by \citet{Remillard06}, it will be classified as an intermediate state. 
We think however that the transition between pure hard states (presence of jets, power-law dominated) and steep power-law states (no apparent jet, disk dominated) should be better highlighted. We therefore use \textit{intermediate} states to represent an identified spectral state, as defined for instance in \citet{HomanBelloni05} and \citet{Nandi12}. We discriminate using only two spectral signatures: the power-law fraction $PLf$, defined as the ratio of the power-law flux to the total flux in the 3--200~keV range, and the power-law index $\Gamma$. This approach leads to the following four spectral states (ignoring quiescence) achieved during an entire typical cycle as reported in Fig.~\ref{fig:Evolution}: 

\begin{figure}[h!]
   \includegraphics[width=1.\linewidth]{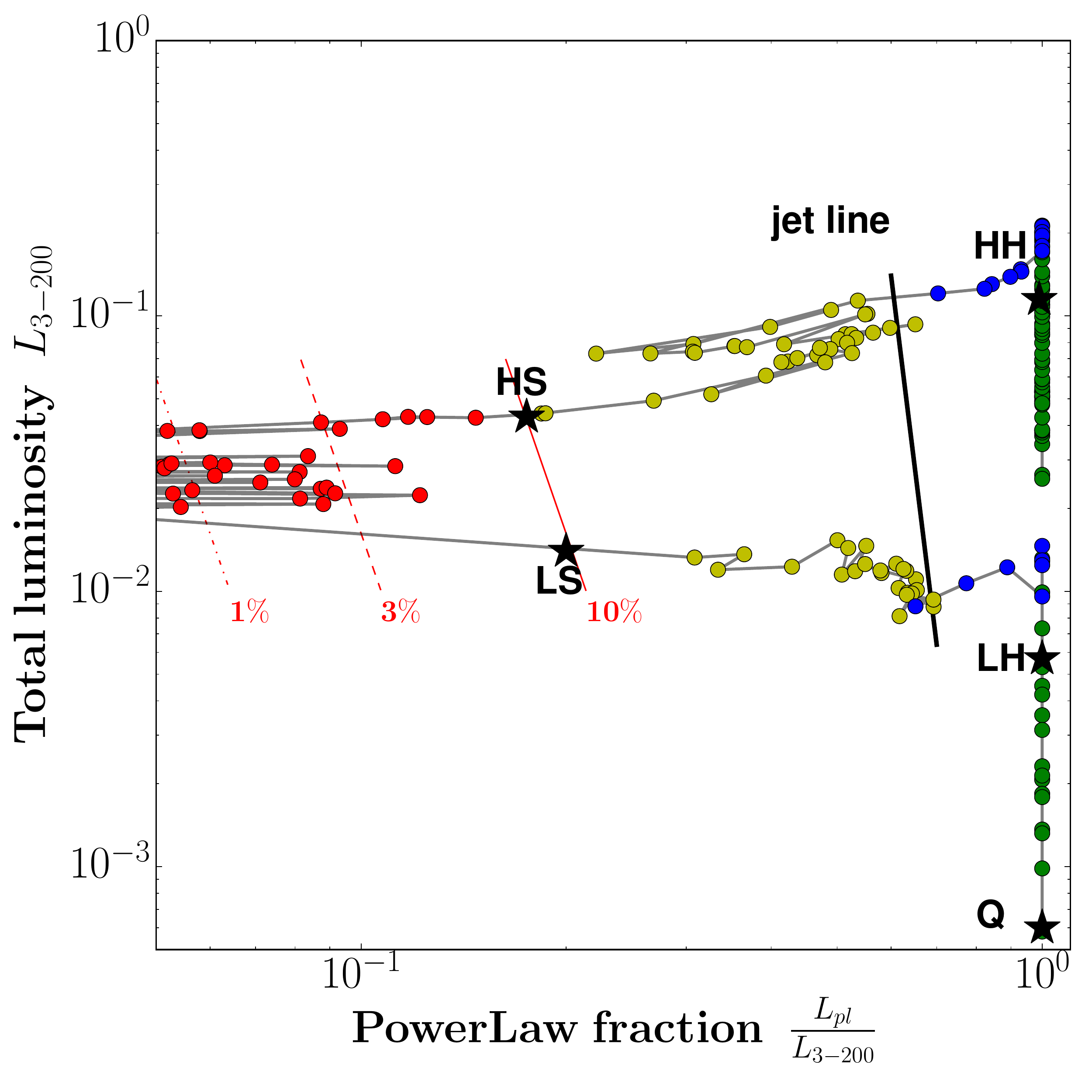}
      \caption{Evolution of \gx during its 2010--2011 outburst in a 3--200~keV DFLD, showing the total $L_{3-200}$ X-ray luminosity (in Eddington units) as function of the power-law fraction. Hard states are displayed in green, hard-intermediate states in blue, soft-intermediate states in yellow, and soft states in red (see text). The five canonical states (Q, LH, HH, HS, LS) defined in paper III are highlighted by the black stars and the observed approximate location of the jet-line is illustrated as a black line. Hard tail levels of $1$, $3$, and $10\%$ are shown in red dotted, dashed, and solid lines, respectively (see text).}
    \label{fig:Evolution}
\end{figure}

\textbf{- Hard states}, in green, are the spectral states where no disk component is detected. They combine quiescent states and more luminous states, as long as the power-law index remains $\Gamma \lesssim 1.8$ with a power-law fraction $PLf = L_{pl}/L_{3-200}=1$ by definition. They appear from day 0 to 85 during the rising phase and between days 400 and 438 during the decaying phase, see Fig.~\ref{fig:Big} top panel, for time evolution.
 
\textbf{- Hard-intermediate states}, in blue, are characterized by a dominant and rather steep power-law, with $1.8 \lesssim \Gamma \lesssim 2.4$ and $PLf > 0.6$. These states, also labeled hard-intermediate in \citet{Nandi12}, are often accompanied by an high-energy cut-off around 50--100~keV at high luminosities \citep{Motta09}. Of course, due to the lack of observations above 40~keV, the cut-off will not be discussed in this work (see discussion Sect.~4.4 in paper~III). They arise from day 86 to 95 in the hard-to-soft transition, then between days 385 and 398 in the soft-to-hard transition.

\textbf{- Soft-intermediate states}, in yellow, are characterized by a disk dominated spectral shape with $0.2 < PLf < 0.6$, accompanied by a reliable steep power-law fit with $\Gamma \simeq 2$--$2.5$. These states were also labeled soft-intermediate in \citet{Nandi12}. They are settled from day 96 to 125 in the hard-to-soft transition and from day 351 to 384 on the way back.

\textbf{- Soft state}, in red, have a disk dominated spectrum with a \HT . This so-called \HT is a steep and faint ($PLf < 0.2$) power-law, with poorly constrained power-law index $\Gamma \in [2,~3]$ (see red portion in Fig.~\ref{fig:Big} bottom-left panel). They occur from day 126 to 317.

Although these states are defined using only two pieces of information: the power-law fraction $PLf = L_{pl}/L_{3-200}$ and the power-law index $\Gamma$, it will be visible that the spectral differences between these states also translate into dynamical differences in the disk evolution. 
This is why extra caution needs to be taken for the soft states, where derived properties also depend on the amplitude of the additional \HT (see paper~III). Since the physical processes responsible for the production of the \HT remain to be investigated \citep[see, e.g.,][]{Galeev79, Gierlinski99}, a constant \HT level of $10\%$ will be used throughout this article. As it clearly appears in Fig.~\ref{fig:Evolution}, this assumption will force us to disregard any observation located on the left hand side of the $10\%$ level line during the soft states. To fully describe these states within the DFLD, one would need to assume modifications in the level of the \HT. Although interesting, this aspect of the problem will not be investigated here, as it affects only a negligible amount of the total energy (see discussions in Sect.~3.2 and 4.1 in paper III).

\subsection{Radio observations} \label{sec:Corbel13}

\begin{figure}[h!]
   \includegraphics[width=1.\linewidth]{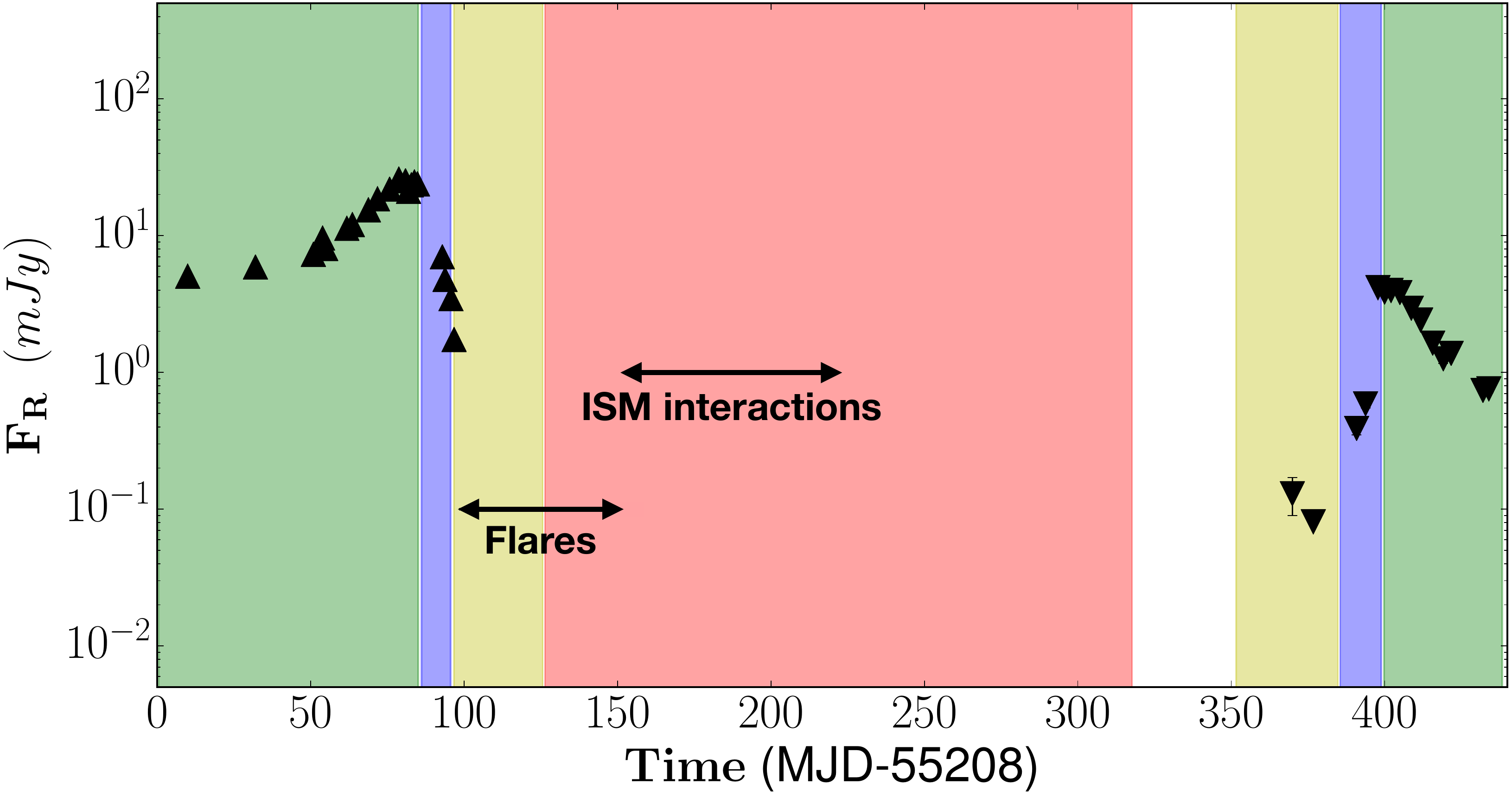}
      \caption{Radio observations at 9~GHz during the 2010--2011 outburst of \gx \citep{2013MNRAS.431L.107C, Corbel13}. Steady jet emission is shown with triangle markers, upper for the rising phase and lower for the decaying phase, all corresponding to proper detections. Double arrows are drawn when radio emission was also observed but has been interpreted as radio flares or interactions with the interstellar medium (Corbel et al. in prep.). The vertical position of the arrows is arbitrary. The background color shades correspond to the X-ray spectral states (see Sect.~\ref{sec:xraydata}). The epoch in white corresponds to a gap in the X-ray coverage due to Sun constraints.}
    \label{fig:Radio}
\end{figure}

During the 2010--2011 outburst, \citet{Corbel13} performed radio observations of \gx with the Australia Telescope Compact Array (ATCA). All radio fluxes likely associated with the presence of steady compact jets obtained during this monitoring are shown in Fig.~\ref{fig:Radio}, with typical statistical errors between $0.01$ and $0.2$~mJy (plotted on Fig.~\ref{fig:Radio}), and systematic errors of typical values $5-10\%$. Two epochs with radio emission are also highlighted, and are interpreted as radio flares or interactions of the ejections with the interstellar medium (Corbel et al., private communication). Since we are mainly focusing on jet diagnostics that can be related to the underlying accretion disk, we will not use the radio fluxes detected during these two periods.

The \textbf{hard states} are always accompanied with radio emission (Fig.~\ref{fig:Radio}, in green). The radio flux increases in the rising phase, and decreases in the decaying phase, implying that radio and X-ray fluxes are likely related. This has indeed been studied in the past, and it has been shown that the radio luminosity $L_R$ at 9~GHz and the X-ray luminosity $L_X$ in the 3--9~keV band follow a universal law $L_R \propto L_X^{0.62}$ \citep{Corbel03, Corbel13}. 
The \textbf{hard-intermediate states} are also characterized by a persistent radio emission (blue). The radio flux varies rapidly, and the spectral slope $F_\nu \propto \nu^\alpha$ changes from the usual $\alpha \in [0,~0.3]$ during the hard states, to a negative slope $\alpha \in [-0.5,~0]$ \citep{2013MNRAS.431L.107C, Corbel13}. This is clear evidence that jets are evolving over time. Due to these spectral index changes during the evolution, we will only consider the radio fluxes in our study rather than the entire jets spectral energy distribution. In the \textbf{soft-intermediate states}, no persistent radio flux is usually observed (yellow). However, multiple radio flares have been observed, mostly during the rising phase, as represented in Fig.~\ref{fig:Radio}. Again, these flares are characterized by a rapidly varying radio emission with negative spectral slopes. In the \textbf{soft state}, no steady radio emission is detected (red). Only few flares and interactions with the interstellar medium (ISM) are observed.

The absence of steady radio emission during both the soft (red) and soft-intermediate (yellow) states gives a significant constraint: we assume that the accretion flow cannot be producing any steady jet in such cases \footnote{Note that some recent studies question whether jets could still be present during the entire evolution, assuming that the detectability of jets is not necessarily straightforward \citep{Drappeau17}. See paper~II, Sect.~5 for previous discussions.}. This requires that the JED is no longer present, or that its size is too tiny to produce any significant radio emitting jet. On the contrary, the presence of steady radio emission during both the hard (green) and hard-intermediate (blue) states is quite easily reproduced with a JED (see Table~1 in paper~III).

\section{Replication of the 2010--2011 cycle} \label{sec:reprod1011}

\subsection{Methodology and caveats}\label{sec:systematics}

As shown in paper III, playing independently with the JED--SAD transition radius $r_J$ and the disk accretion rate $\dot{m}_{in}$ allows us to compute spectra and to reproduce thereby any particular position within the DFLD. We thus perform a large set of simulations with $r_J \in [r_{in}=2,~100]$ and $\dot{m}_{in} \in [10^{-3}, ~10]$, computing for each pair \pair the thermal balance of the hybrid disk configuration, and its associated theoretical global spectrum. We then fake data and fit each resulting spectrum using the \xspec fitting procedure detailed in paper~III. We use the same\footnote{Despite the fact that we used \textsc{wabs} as absorption model in paper III, we now use \textsc{phabs}. Note that in the energy bands covered by PCA, the differences are barely detectable.} spectral model components (power-law and disk) as those used for the spectral analysis of the real observations \citep[see Sect.~\ref{sec:xraydata} and][]{Clavel16}. As a consequence, the theoretical parameters can be directly compared to the observational parameters, namely: the 3--200~keV luminosity $L_{3-200}$, the power-law flux $L_{pl}$ in the same energy range, the power-law fraction $PLf = L_{pl} / L_{3-200}$, and the power-law photon index $\Gamma$. 

However, there are several differences between the spectra obtained from observations (hereafter, observational spectra) and those obtained from our theoretical model (hereafter, theoretical spectra), that could induce systematic effects in this comparison. 

\begin{figure*}[t]
  \includegraphics[width=1.0\linewidth]{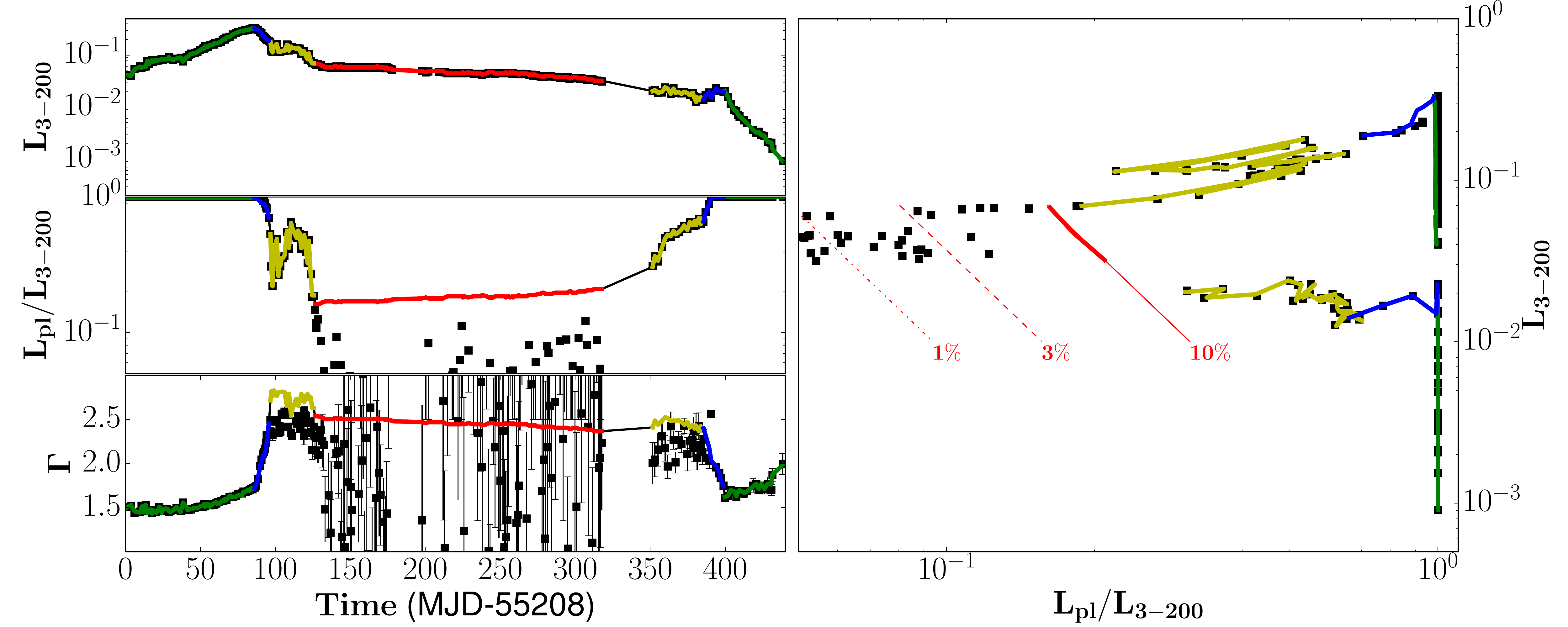}
   \caption{Results of fitting procedure A applied to the 2010--2011 outburst of \gx . The black markers are fits taken from \citet{Clavel16} reported with their error-bars when reliable, while color lines display our results: green, blue, yellow and red for hard, hard-intermediate, soft-intermediate, and soft states, respectively (see Sect.~\ref{sec:GX}). Left, from top to bottom: the 3--200~keV total luminosity $L_{3-200}$ (in Eddington units), the power-law luminosity fraction $PLf = L_{pl}/L_{3-200}$ and the power-law index $\Gamma$. Right: the DFLD. The \HT proxy used was frozen to $10\%$, but $1\%$ and $3\%$ proxies are also illustrated as red dot-dashed and dashed lines (see paper III).}
  \label{fig:Big}
\end{figure*}

First, theoretical spectra are created using both \textit{RXTE}/PCA and HEXTE instrumental responses (for $1$~ks exposures) and thus cover the full $3-200$~keV range. This larger energy range allows better constraints on the spectral shape of the theoretical spectra. In particular, the high-energy cut-off detected in part of the associated fits cannot be constrained in the observational spectra, limited to energies below 40~keV. In order to correct for this difference, these high-energy cut-offs were ignored when computing the theoretical luminosities ($L_{\rm pl}$ and $L_{\rm 3-200}$). The quality of the theoretical spectra also allows us to detect faint low-temperature disk components which would not be considered significant in the observational spectra. However, this difference only affects a few points in the DFLD (close to the high hard state) and the shifts induced are small enough to be neglected (see, e.g., Fig.~\ref{fig:Big}). 

Second and most importantly: in addition to the power-law and disk components described above, a 'reflection' component (traced by a strong emission line at $\sim6.5$~keV) is also detected in most observational spectra. The reflection process is not currently implemented in our theoretical model, so the theoretical spectra do not include any reflection component. Therefore, the model chosen to account for the reflection signal in the observational spectra could induce systematic errors, mostly in the power-law parameters, and it is important to quantify them. 
To account for the reflection component, \citet{Clavel16} selected an ad hoc model composed of a Gaussian emission line at $6.5$~keV and, when needed, of a smeared absorption edge at $7.1$~keV. These components mimic the shape of the reflection features without accounting for the full complexity of the problem \citep[see, e.g.,][and references therein]{2014ApJ...782...76G, 2014MNRAS.444L.100D}. Indeed, individual \textit{RXTE}/PCA observations have an energy range, a spectral resolution, and an exposure time that prevent more complex reflection models to be fitted on the corresponding spectra. The comparison between the spectral parameters obtained with our ad hoc model and those derived using self-consistent reflection models was therefore done after merging several observations together to improve the available statistics \citep[see][for an example of such spectral analysis for \gx in the hard state]{Garcia15}. 

Our investigation confirms that among all parameters used in the present work, the power-law photon index $\Gamma$ is the most impacted. In particular, the use of our ad-hoc model tends to underestimate the value of $\Gamma$ with discrepancies up to $\Delta\Gamma\lesssim0.2$. This is especially true in the hard state where such a shift would reduce the difference between theory and observations that was introduced by taking into account the radio constraints (see Fig.~\ref{fig:ProcB_res}, 4th panel on the left). Due to the extrapolation of the observed luminosity from 3--40~keV to 3--200~keV, such an increase of $\Gamma$ would also induce a decrease of $L_{\rm pl}$ by at most $20\%$. In addition, merged spectra provide enough statistics to better constrain the power-law component present in the softer states (yellow and red in, e.g., Fig.~\ref{fig:Big}), also leading to shifts in the value of the photon index $\Gamma$. Once again, this shift between the parameter obtained from grouped spectra and the median of parameters from individual fits are no larger than $\Delta\Gamma\lesssim0.2$, inducing variation on $L_{\rm pl}$ of at most $15\%$.

Having these caveats in mind, we can turn to the comparison between theoretical spectra and observations.

\subsection{Fitting procedure A: X-rays only} \label{sec:Xray}

For each observation, we search for the best pair \pair that minimizes the function
\begin{equation}
\zeta_X =  \frac{ | \text{log} \left[ L_{3-200} / L_{3-200}^{obs} \right] | }{\alpha_{flux}} +  \frac{ | \text{log} \left[ PLf / PLf_{obs} \right] | }{\alpha_{PLf}} + \frac{ | \Gamma - \Gamma^{obs} | }{\alpha_{\Gamma}}
\label{eq:Xray}
\end{equation}
\noindent where $L_{3-200}^{obs}$, $PLf^{obs}$, and $\Gamma^{obs}$ are the values derived from \cite{Clavel16}. The coefficients $\alpha_{flux}$, $\alpha_{PLf}$, and $\alpha_{\Gamma}$ are arbitrary weights associated to each one of the three constraints considered. The relative importance of the flux and power-law fraction are comparable, so we choose $\alpha_{flux} = \alpha_{PLf} = 1$.  More caution needs to be taken on the weight $\alpha_{\Gamma}$ put on the spectral index $\Gamma$. Indeed, while quite constraining during the hard states, the value of $\Gamma$ becomes unreliable during the soft state. We thus choose $\alpha_{\Gamma}$ to be a function of the power-law fraction, namely $\alpha_{\Gamma}  = 2 - 6~ \text{log}_{10} (PLf)$. Note that the lowest value of the power-law fraction is reached in the soft state and depend on the (chosen) \HT level, namely 0.1 here. As a result, $\alpha_{\Gamma}$ varies from $2$ in hard states to $8$ in the soft state. Although empirical, this choice of $\zeta_X$ has revealed very effective in providing the best pair of parameters \pair for a given spectral shape. Fitting the entire cycle using this procedure takes only a few seconds.

\subsubsection{Results: DFLD and spectra} 
\label{sec:spectre}

The best fits obtained using procedure A are shown in Fig.~\ref{fig:Big}. Observations are displayed in black and each reproduced state appears with its corresponding color code, as in Fig.~\ref{fig:Evolution}. Since the canonical spectral states can be easily reproduced (paper~III) it is not surprising that the whole cycle evolution in the DFLD can also be successfully recovered. This is however the first time that such a study is presented. While the total X-ray flux is satisfactorily reproduced (Fig.~\ref{fig:Big}, top-left panel), additional comments are necessary for the evolution of both the power-law fraction $PLf$ and the spectral index $\Gamma$.

Most of the evolution of the power-law fraction (middle left panel) is nicely recovered except for two epochs: one around days 80-90 (blue) and the other during the soft states (red). The zone around days 80-90 corresponds to the upper transition from the rising hard state to the high hard-intermediate state. While no disk is detected in the X-ray observation, our model often includes a weak disk component, decreasing the power-law fraction by $1-2\%$ in the worst cases (see upper transition in the DFLD). The presence of weak disks in luminous hard states remains debated \citep[see, e.g.,][]{Tomsick08}, but may be solved by missions with soft X-ray sensitivity like NICER. For the present work, \textit{RXTE} data alone provide no constraints below 3~keV and investigating the presence of such a component is therefore not possible with our data set. Note that disk detection in hard states with \textit{RXTE} are often considered as unrealistic due to the parameters derived, see for example \citet{Nandi12} or \citet{Clavel16}. In the second ill-behaved portion, associated with the soft state (in red, days 126 to 317), our $PLf$ is larger than the observed values. This is a natural consequence of our choice of a $10\%$ level for the \HT . As discussed earlier, this bias could be easily accounted for by allowing the \HT level to vary in time, down to $1\%$ or lower (Fig.~\ref{fig:Big}, right panel). 

The time evolution of the power-law index $\Gamma$ is represented in Fig.~\ref{fig:Big}, bottom left panel. Our findings are in agreement with the observations whenever they are reliable (away from the soft state in red). The only noticeable difference lies in the soft-intermediate stages (days 90-120 and 350-380), where the power-law is slightly steeper than observed, namely $\Gamma = 2.4$--$2.6$ compared to $\Gamma_{obs} = 2.1$--$2.5$. But this discrepancy is only of the order of 0.2--0.3, a range consistent with the largest systematic errors introduced by reflection component issues (see Sect.~\ref{sec:systematics}). Choosing self-consistent models to account for this component would presumably lead to slightly larger values of $\Gamma_{obs}$, decreasing the difference with our $\Gamma$. To be fully exhaustive, this could also be accounted for by varying model parameters frozen in papers II and III (e.g., accretion speed, illumination processes).

\begin{figure}[t!]
  \includegraphics[width=1.0\linewidth]{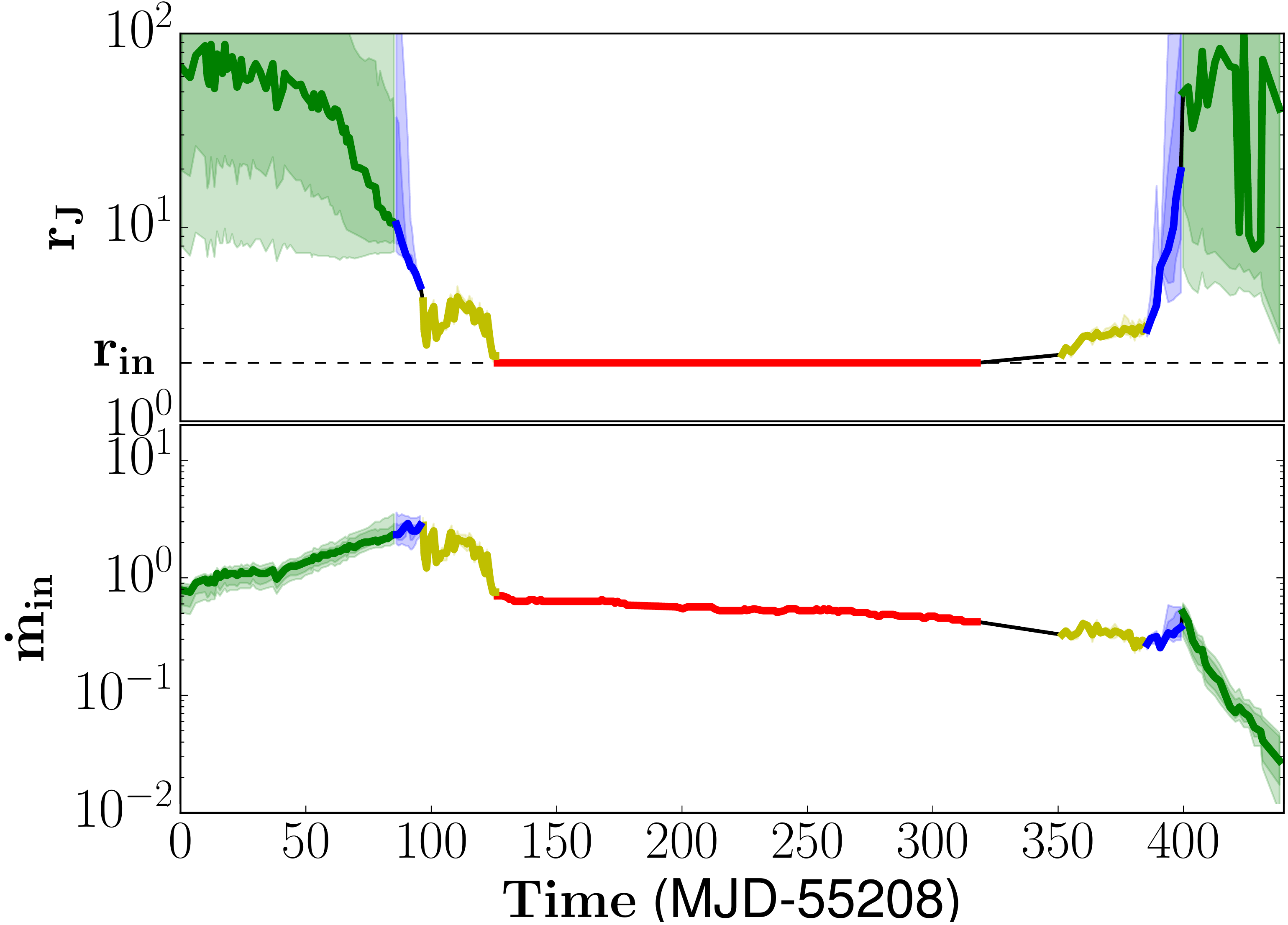}
   \caption{Time evolution of \rj (top) and \mdotin (down) associated with the best $\zeta_X$ defined by Eq.~(\ref{eq:Xray}). The color code is the same as previously. The transparent colored areas correspond to the confidence intervals of $5\%$ and $10\%$ error margin (see text).}
  \label{fig:RjAndMdot}
\end{figure}

Taking all these elements into consideration, it appears that both individual spectra and global DFLD evolution are quite nicely reproduced by the model. The evolutionary track in the DFLD has been obtained by finding the best fits \pair according to the minimum $\zeta_X$ at each time (Eq.~\ref{eq:Xray}). The corresponding evolutionary curves $r_J(t)$ and $\dot m_{in}(t)$ are shown in Fig.~\ref{fig:RjAndMdot}. Each color represents the spectral state of \gx while the surrounding colored areas represent confidence intervals. The most transparent area corresponds to the pairs of \rj and \mdotin for which $\zeta_X$ is within the $10\%$ error margin of the fits, namely $\zeta_X < 1.10 \zeta_{X,\text{best}}$. The least transparent one corresponds to a $5\%$ error margin such that $\zeta_X < 1.05 \zeta_{X,\text{best}}$. It is worth nothing that the accretion rate is very well constrained during the whole evolution. On the contrary, the transition radius $r_J$ has good constraints only during the hard-intermediate (blue) and soft-intermediate (yellow) states, while during the hard states, large differences in $r_J$ are barely visible in the resulting spectra. 

The existence of a q-shaped evolutionary track in the DFLD, attributed to an hysteresis, is here replicated by playing with two (apparently) independent parameters, the transition radius \rj between the inner JED and the outer SAD and the disk accretion rate $\dot m_{in}$. The two time-series shown in Fig.~\ref{fig:RjAndMdot} are smooth and follow qualitatively the expected behavior of hybrid disk models \citep[see, \eg ,][]{Esin97, Done07,Ferreira06, Petrucci08, BegelmanArmitage14, Kylafis15}. While it is quite natural to expect some convexe curve for $\dot m_{in}(t)$ (or $\dot{m}(t)$ at a given radius) during an outburst, the behavior of \rj remains a mystery. During the quiescent phase, most of the inner regions of the disk need to be in a JED accreting mode. Then, as the disk accretion rate increases, there must be an outside-in decrease of the transition radius, leading to the diminishing until final disappearance of the inner JED when \rj reaches $r_{in}$. Once the binary system reaches the soft portion of its evolution, it remains so until the decrease in $\dot m_{in}$ leads to an inside-out rebuilding of the inner JED when \rj increases again. 

Our work reveals what would be required in order to explain the 2010--2011 outburst of \gx without explaining the reasons for it. What triggers the growth and decrease of the JED requires dynamical investigations that are beyond the scope of the present study. We note however that, within our paradigm, this must be related to the strength and radial distribution of the vertical magnetic field embedding the disk, as proposed in paper~I.

\subsubsection{Results: predicted radio light curve} 
\label{sec:dynamics}

\begin{figure}[h!]
  \includegraphics[width=1.0\linewidth]{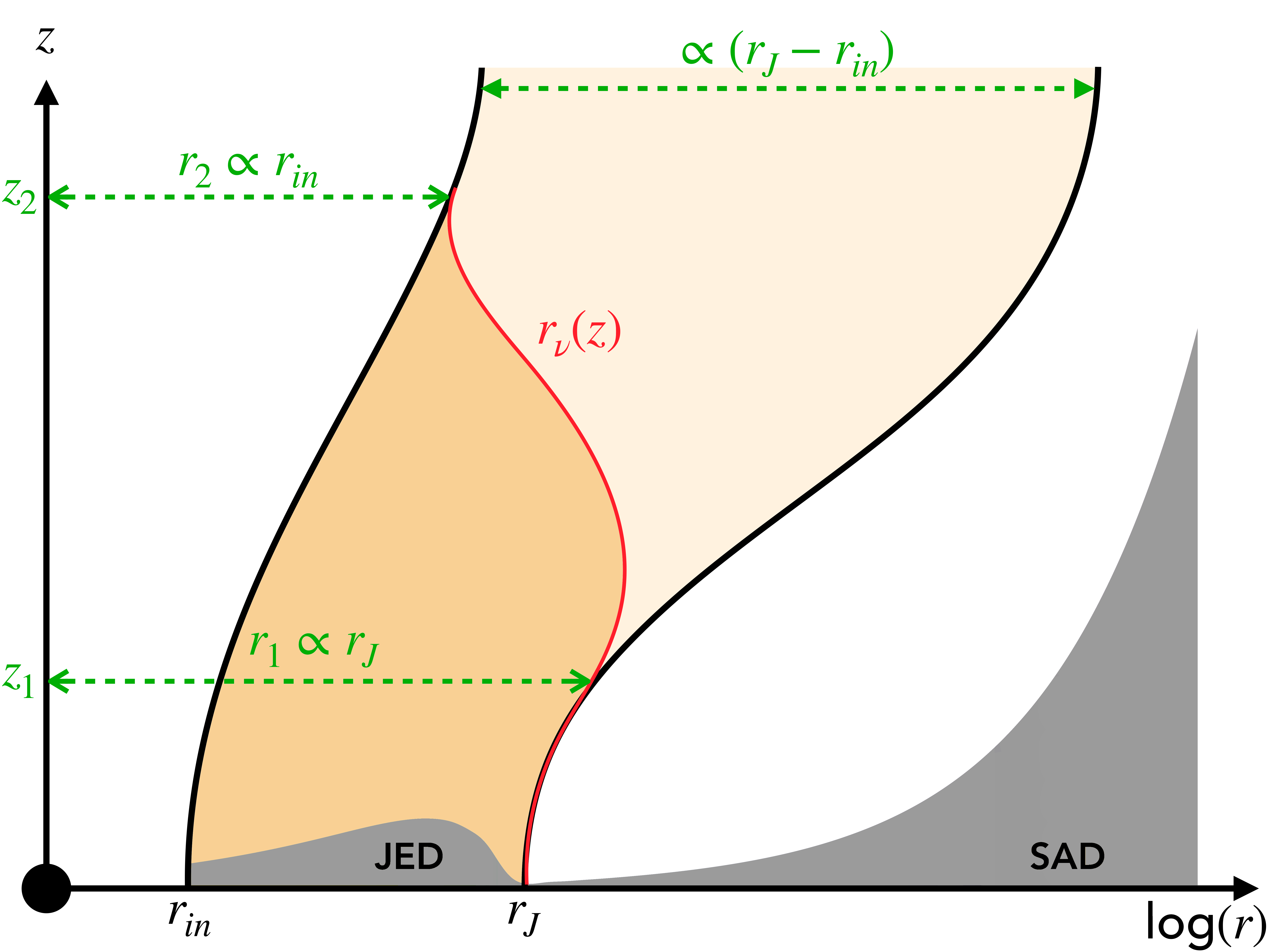}
   \caption{Schematic view of the magnetized accretion-ejection structure, as a function of radius $r$ and altitude $z$. The extent of the JED and SAD is shown in grey (vertical heights were computed from our model, see paper~III), while the jet launched from the JED is shown in orange (sketch of what its geometry could look like). The photosphere $r_\nu(z)$ at a given frequency $\nu$ is shown in red, splitting the jet in an inner optically thick (dark-orange) and outer optically thin (light-orange) regions. Within our simple approach, each field line widens in a self-similar way so that, at a given altitude $z$, a field line anchored at $r_{in}$ has achieved a radius $r(z)= W r_{in}$, while that anchored at $r_J$ is $W r_J$, with the same asymptotic widening factor $W$.}     
  \label{fig:Rad}
\end{figure}

This section aims at revising previous estimates of the radio flux emitted by a given accretion flow \citep[see for example][]{BK79,Heinz03}. In our estimates, we neglect any radiative contribution from the central Blandford $\&$ Znajek jet core \citep[][see introduction of paper II]{BZ77}. Furthermore, by considering only radio fluxes no assumptions\footnote{Note that such an assumption is justified by the existing radio/X-ray correlation \citep{Corbel13}, despite variations of $\alpha$ during the evolution.} on the spectral index of jets have to be made (see paper III, Appendix A).
Radio emission is assumed to be self-absorbed synchrotron emission from a non-thermal power-law particle distribution with an exponent $p=2$. The local magnetic field in the jet is assumed to follow that of the disk in a self-similar fashion. The magnetic field in the JED is known since it depends only on the local disk accretion rate. All usual uncertainties such as details of particle acceleration, jet collimation, Doppler beaming, or inclination effects can be hidden within a common normalization factor. Although legitimate, a more precise description of these processes is far beyond the scope of this paper. There are still, however, uncertainties related to the shape of the photosphere at the given radio frequency (e.g., $\nu_R = 9$~GHz), as illustrated in Fig.~(\ref{fig:Rad}).

Since the JED has a finite radial extent, the whole jet is itself limited in radius with a total width proportional to $r_J-r_{in}$. At a frequency $\nu_R$, the whole jet is thus optically thick up to an altitude $z_1$, whereas it becomes fully optically thin beyond $z_2$. Between these two altitudes, the photosphere has a shape following $r_\nu(z)$ whose precise description can only be known once a full 2D  (not self-similar) jet model will be achieved. This is still a pending issue. Therefore, an approximation for this photosphere is required to estimate the flux emitted in the radio band. 

In paper~III, it was implicitly assumed that the dominant radio emission would be emitted around $z_1$, leading to the following expression for the flux 
\begin{equation}
F_{R_1} = \tilde f_{R_1} ~ \dot{m}_{in}^{17/12} r_J \left ( r_J - r_{in} \right )^{5/6} ~ F_{Edd}
\label{eq:Fradio1}
\end{equation}
where $F_{Edd}= L_{Edd}/(\nu_R 4\pi d^2)$ is the Eddington flux at $\nu_R$ received at a distance $d$. As expected, the radio flux $F_{R_1}$ is not only a function of the disk accretion rate \mdotin , but also a function of the JED radial extent [$r_{in}$, $r_J$]. All usual uncertainties are incorporated within the normalization factor $\tilde{f}_{R_1}$, which can be tuned to fit the observed fluxes. We refer the reader to Appendix A in paper III for more informations on the derivation of this equation~(\ref{eq:Fradio1}).

However, in the limit of a large JED extension, the radio flux scales as $F_{R_1} \propto r_J^{11/6}$ with this formula. Although increasing the emitting volume (or, alternatively, the photosphere surface scaling as $r_{J} r_{in}$) might lead to an increase of the radio flux, it seems doubtful that the flux would increase almost like $r_J^2$. Indeed, radio is due to self-absorbed synchrotron emission that depends on the local magnetic field strength. Now, while there is currently no consensus on the magnetic field distribution in jets, they are usually described with a central core of almost constant field, surrounded by a steeply decreasing magnetic field structure \citep[see, \eg ,][]{2015MNRAS.447.2726N}. As a consequence, we expect a swift transition of the photosphere $r_\nu(z)$, going from the external jet radius (at $z_1$) to the internal jet interface (at $z_2$), where magnetic fields are stronger. Radio jet emission could thus be actually dominated by a zone of surface $\propto r_{in}^2$ around $z_2$. Within this other extreme limit, one derives the following expression for the radio flux  
\begin{equation}
F_{R_2} = \tilde f_{R_2} ~ \dot{m}_{in}^{17/12} r_{in} \left ( r_J - r_{in} \right )^{5/6} ~ F_{Edd}
\label{eq:Fradio2}
\end{equation}
This new expression is better behaved at large JED radial extent and remains consistent with a drop in the radio flux when the JED disappears ($r_J \rightarrow r_{in}$). Clearly, it is impossible to go further without a proper 2D calculation of the jet dynamics. But, within our current level of approximation for the jet radio emission, the above two expressions provide two reasonable flux limits. 

\begin{figure}[h!]
  \includegraphics[width=1.0\linewidth]{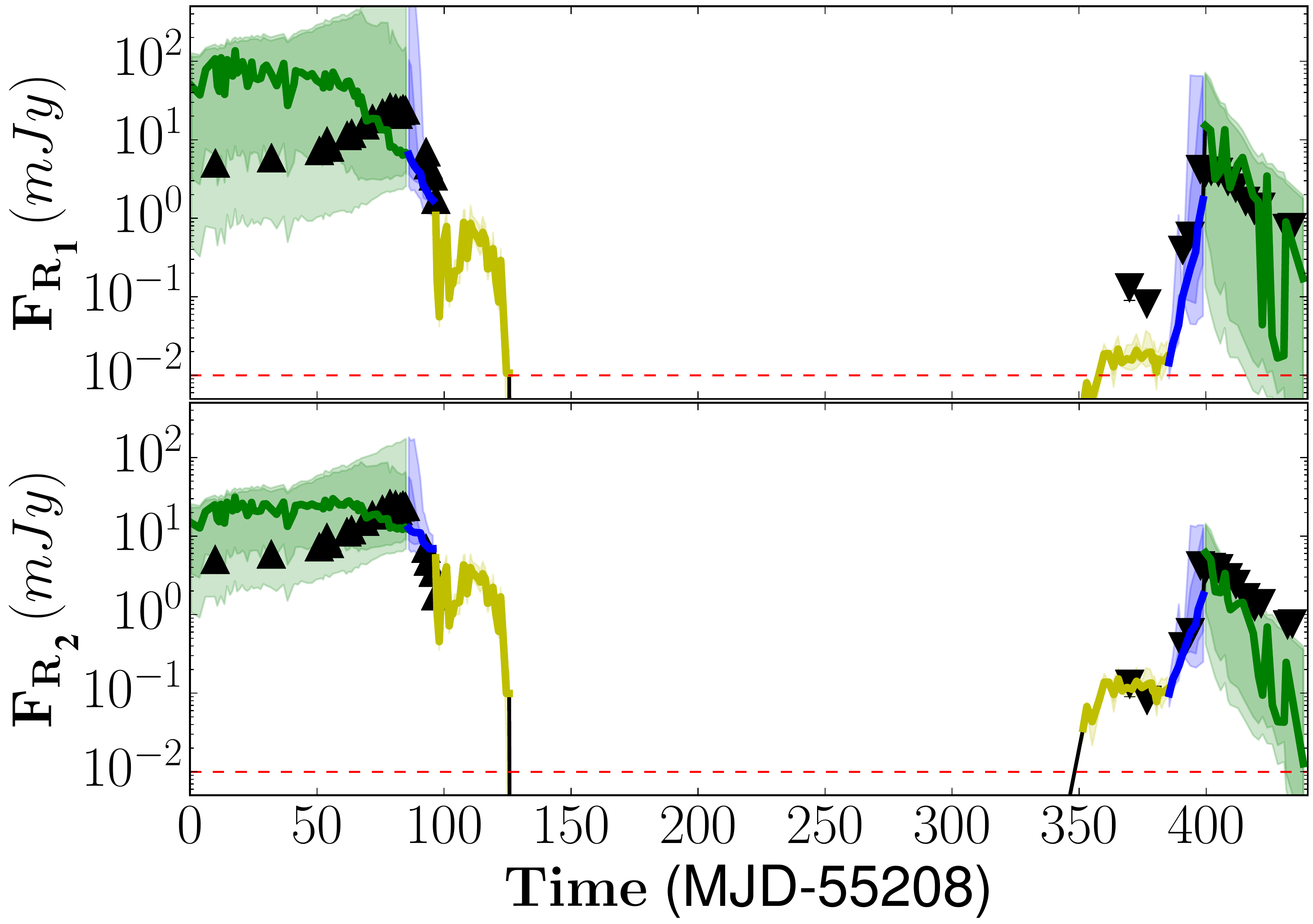}
   \caption{Radio light curves at 9~GHz in mJy predicted using Eq.~(\ref{eq:Fradio1}) (top) and Eq.~(\ref{eq:Fradio2}) (bottom). A red dashed line at $10^{-2}$~mJy illustrates a typical detection limit. Markers correspond to observed radio fluxes associated with steady jets (see Fig.~\ref{fig:Radio}.}
  \label{fig:ProcA_Fr}
\end{figure}

Figure~\ref{fig:ProcA_Fr} displays the \citet{Corbel13} radio observations (black), and the two predicted radio light curves using the parameters found on X-ray spectral modeling only. The two normalization factors, $\tilde{f}_{R_1} = 1.05 \times 10^{-10}$ and $\tilde{f}_{R_2} = 4.5 \times 10^{-10}$, are assumed to remain constant along the full cycle and were chosen to best fit radio observations at $\nu_{R}=9$~GHz. 
Strikingly, the global trends follow reasonably well the observations, with a net radio decrease (resp. increase) near the switch-off (resp. -on) phase in both cases. Clearly, the level of radio emission is always adjustable via the renormalisation factor $\tilde{f}_{R}$. But what is noteworthy here is the shape of the light curve itself, since again it is obtained using only constraints ($r_J, \dot{m}_{in}$) derived from X-ray spectra. This result is promising since the reproduction of the radio fluxes was not a requirement of our fitting procedure. 

There are, however, some quantitative differences that deserve some attention. The theoretical radio flux is overestimated in the rising hard state by up to two orders of magnitude (upper triangles, days 0--80). This is especially the case for $F_{R_1}$, due to the higher dependence on the transition radius $r_J$. The same problem arises in the decaying phase, when the flux is underestimated at first (days 360--380) and overestimated later in the hard state (days 400--450). Note however that most of these points belong to the $5\%$ error bar regions (or confidence intervals) for $r_J$. The fact that observed points always lie within these regions means that a reasonable radio light curve can be built using values of $r_J(t)$ derived using the X-rays, even at epochs when X-rays are barely sensitive to $r_J$ (largest error bars). Moreover, when the uncertainties in $r_J$ decrease (blue zone) the model adequacy for the radio increases significantly. Since error bars in radio are quite small, explaining the radio emission during the hard states would only require to change $r_J$ accordingly, with no significant loss of information in the X-rays. In practice, this discussion advocates for the inclusion of radio luminosities within the fitting procedure, as described in Sect.~\ref{sec:Xray+R}.

A different kind of discrepancy between the radio emission models and data can be seen during the yellow soft-intermediate phases. In the high phase (days 100--130) the transition radius has not yet decreased down to $r_{in}$ (corresponding to a zero-extent jet) and, according to both our formulae, radio emission should be produced and detected. However, no detection of steady jets is reported, only flares (see \fig~\ref{fig:Radio}). In the low phase (days 350--380), radio emission is detected at a level sometimes larger than predicted by the model (especially for $F_{R_1}$). Our simple jet emission model seems thus to introduce significant errors during these phases. However, they correspond to $r_J$ varying between $r_{in}$ and $2 r_{in}$ (\fig~\ref{fig:RjAndMdot}). Now, the jet emission model assumes that the radio emission properties scale with the JED radial extent. It is doubtful that this would still hold when $r_J \rightarrow r_{in}$. Indeed, jet collimation properties (and possibly particle acceleration efficiency) may undergo significant changes when the jet launching zone becomes a point-like source. We should therefore not expect too much from our radio emission model during these states. 

To summarize, our simple radio emission model does seem to predict the global trend for the observed radio light curve, for both formulae. This was a by-product of our JED-SAD formalism and not a requirement of the model. However, not only the theoretical dependency of $F_{R_1}$ with $r_J$ appears dubious, but it is also less effective in reproducing the observed radio light curve. This is the reason why, in the following, we will only focus on $F_{R_2}$.

\section{Replication of the 2010--2011 cycle including radio constraints}
\label{sec:Xray+R}

\subsection{Fitting procedure B} 

Hereafter, the radio flux is described by Eq.~(\ref{eq:Fradio2}), and $F_{R_2}$ and $\tilde f_{R_2}$ will now be referred to as $F_{R}$ and $\tilde f_{R}$. Being very sensitive to the transition radius $r_J$, this radio flux can thus be used along with the X-rays to constrain the disk dynamical state. Hence, for each observation at any given time, we now search for the best pair \pair that minimizes the new function
\begin{equation}
\zeta_{X+R} =  \zeta_X + \frac{ | \text{log} \left[ F_R / F_R^{obs} \right] | }{\alpha_{R}} \label{eq:Xray+Radio}
\end{equation}
\noindent where $\zeta_X$ is defined in Eq.~(\ref{eq:Xray}), $F_R$ is the predicted radio luminosity computed above, $F_R^{obs}$ the observed luminosity, and $\alpha_{R} = 5$ the radio weight chosen for best results. Note however that the addition of the radio constraints is not straightforward for the three following reasons:

(1) We only have 35 radio observations across the $\sim 450$~days of the observed cycle. Since no radio flux associated with compact jets has been observed during most of the soft-intermediate states (yellow) and during the soft state (red), no radio constraint will be imposed during these states (see Fig.~\ref{fig:Radio}). We thus use the former $\zeta_X$ (Eq.~\ref{eq:Xray}) in these cases.

(2) When they exist (hard and hard-intermediate states), radio observations often do not exactly match with the date of X-ray observations. However, during these states, thanks to the high density of radio observations, we were able to interpolate radio observations and to associate to each X-ray observation an expected radio flux.

(3) Theoretical radio fluxes $F_R$ are computed with Eq.~(\ref{eq:Fradio2}) using the same normalization factor for the entire cycle. This is justified as long as some self-similar process is involved throughout the cycle. Note that, contrary to the fitting procedure A, including the radio introduces a global constraint through this common value $\tilde{f}_R$. Within procedure A, the radio light curve could be shifted vertically (in flux) by changing the value of $\tilde{f}_R$ without modifying either $r_J$ or $\dot m_{in}$. In procedure B, the value of $\tilde{f}_R$ is included in each individual fit and thus influences the convergence to the best \pair parameter set. Our best value is $\tilde{f}_{R} = 2.5 \times 10^{-10}$ (remember that $r_{in}=2$ here).

\begin{figure*}[t!]
   \includegraphics[width=1.0\linewidth]{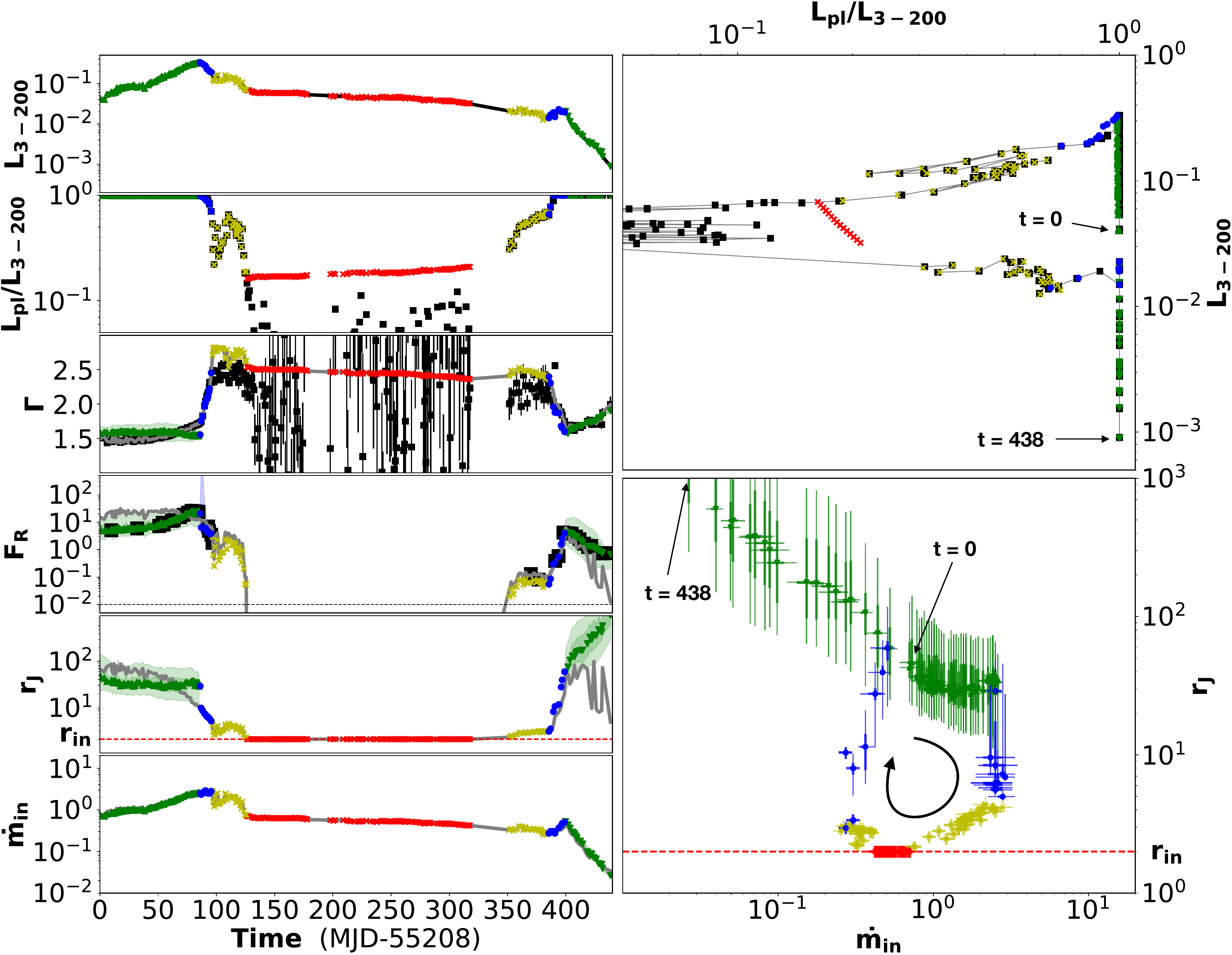}
   \caption{The 2010--2011 outburst of \gx , derived using the fitting procedure B that takes into account X-rays and radio constraints. On the left, from top to bottom, light curves of: 3--200~keV flux (in Eddington units), power-law fraction, power-law index, radio flux $F_R$ (in mJy), transition radius \rj , and accretion rate \mdotin . Light curves of the same quantities, but obtained using procedure A (Fig.~\ref{fig:Big}), are also reported with grey lines for comparison. On the right, evolutionary tracks in the associated DFLD (top) and in a \rj vs \mdotin diagram (bottom), in which the starting and ending positions are indicated. The color code is the same as previously: green triangles for hard state, blue circles for hard-intermediate, yellow crosses for soft-intermediate and red crosses for soft. Different symbols were used for the two phases of the hard state to better disentangle them: Upper-triangles are used for the rising phase and lower-triangles for the decaying phase. Error-bars for the $5\%$ and $10\%$ confidence intervals are also displayed on the bottom-right panel.}
  \label{fig:ProcB_res}
\end{figure*}

\subsection{Results}

Figure~\ref{fig:ProcB_res} shows the results of the fitting procedure B using both X-rays and radio fluxes. We use the same color-code as before: green for hard, blue for hard-intermediate, yellow for soft-intermediate, and red for soft states. 

In comparison to procedure A, this new procedure introduces modifications to the transition radius \rj and marginally to the disk accretion rate \mdotin . This leads to significant modifications to the radio flux $F_R$ and the spectral index $\Gamma$. To better see these changes, we over-plotted the results from procedure A in grey lines in Fig.~\ref{fig:ProcB_res}. As expected, the global X-ray flux and power-law fraction are not altered by the addition of radio constraints. Their evolution as well as the associated DFLD remain the same as in Fig.~\ref{fig:Big}.
%
The strongest quantitative modifications to the global evolution of \rj and \mdotin occur in the hard states (Fig.~\ref{fig:ProcB_res}, two bottom left panels). Instead of monotonically decreasing in time (as with procedure A), the transition radius now remains roughly stable around $r_J \simeq 20$, i.e. $r_J \simeq 10 r_{in}$, within most\footnote{Note that this is consistent with previous calculations using \citet{Heinz03} calculations, where \rj was not taken into account in the model.} of the hard state evolution and especially in the rising phase between days 0 and 80. This smaller transition radius also leads to a slightly smaller accretion rate, due to the higher radiative efficiency of the SAD (paper~III).

Spectrally, the reproduction of the power-law index $\Gamma$ is slightly degraded in the rising hard states (days 0--80). The other portions of the outburst are not affected, due to the absence of radio constraint, and their reproduction remains satisfactory. In the hard states, the addition of a radio constraint forces a smaller transition radius than before (Fig.~\ref{fig:ProcB_res}, left column, 5th panel). This results in spectra harder than those observed, with $\Gamma \simeq 1.6-1.7$ as compared to $\Gamma_{obs} \simeq 1.5-1.6$ (3rd panel). 
As argued in Sect.~\ref{sec:systematics}, this discrepancy could be easily compensated by better taking into account the reflection component. Moreover, although our model deals with the illumination by the inner JED of the outer soft SAD photons, it does so in quite a simplified way using a simple geometrical factor $\omega$. As shown for instance in Fig.~2 from paper~III, a slight modification of $\omega$ could lead to variations up to 0.4 in $\Gamma$. The systematic error induced by these two simplifications has the right order of magnitude to explain the discrepancies seen in Fig.~\ref{fig:ProcB_res}. Therefore, the model is consistent with observations for the whole cycle.
 
Finally, and by construction, the addition of the radio fluxes within the fitting procedure allowed to much better reproduce the radio light curve (Fig.~\ref{fig:ProcB_res}, left column, 4th panel). Both the rising and decaying radio emissions during the hard states (green) now match almost perfectly. The hard-intermediate states (blue), during which a swift decrease/increase in radio luminosity occurs, are also finely recovered. With no surprise, as with procedure A, the behavior of our jet emission model during the soft-intermediate states (yellow) has room for improvement: too much emission is predicted in high states. This issue is already discussed in Sect.~\ref{sec:dynamics}. 

The bottom right panel in Fig.~\ref{fig:ProcB_res} shows the co-evolution $r_J (\dot m_{in})$ of the disk accretion rate and the JED-SAD transition radius. The overall behavior follows qualitatively what was expected from previous theoretical works from, for example, \citet{Meyer05} or \citet{Kylafis15}. Note that these were only illustrations, whereas our $r_J (\dot m_{in})$ curve relies on complex modeling and fits performed on multi-wavelength data. Such a representation suggests that the disk accretion rate is the only control parameter that leads to the observed evolutionary track of XrBs. This is not a conclusion that can be drawn yet. Indeed, $r_J$ depends on the local disk magnetization (hence a function of the local magnetic field and disk column density) while $\dot m_{in}$ depends on both the dominant torque and column density. There are thus two local disk quantities that come into play: the magnetic field and the column density; and the way the magnetic field evolves in accretion disks is still a pending question. One possible way to tackle this issue would be to see how generic the resulting $r_J (\dot m_{in})$ curve is, for instance by deriving it for all cycles of a given object as well as for different objects.

\subsection{Open questions}

Before concluding, a few open questions need to be discussed.

First, in our view jets are mainly emitted directly from the disk, i.e. the inner JED portion, and we only computed the radio emission from that component. Since the global scenario from paper~I relies on the existence of a large scale vertical magnetic field, jets emitted from the rotating black hole ergosphere are also probably present \citep{BZ77, Tchekhovskoy11, McKinney12}. We did not take this dynamical component into account and doing so could potentially lead to some quantitative changes in the total radio emission. This deserves further investigation. Moreover, multiple radiative processes concerning the two jets have been ignored. We aim here at giving a first order estimate of the jets contribution, but a more precise description of the jets radiative emission signature would lead to a more realistic picture. One could for example imagine including effects like jet collimation or inclination. This is a work that will be done by coupling our code with a jet spectral code ISHEM \citep{2014MNRAS.443..299M, 2019MNRAS.482.2447P}, and thus also further addressing the possibility of dark jets \citep{Drappeau17}. Additionnally, a careful reader will note that our face-on disk is inconsistent with our edge-on estimates of the jets emission: this issue will be dealt with in the future.

Second, another aspect that has been disregarded in our study so far is the possible production of winds. In X-ray binaries, winds have mainly been detected in the soft state \citep{2002ApJ...567.1102L, Ponti12}, but very recent work suggests the discovery of such winds in the hard state, challenging this usual and historical view \citep[see for example][]{2016ApJ...830L...5H, 2018Natur.554...69T, 2018MNRAS.481.2646M}. Unless, of course, the mass loss rate in the wind is huge, the presence of such winds in the soft state would not have an important impact on the disk structure, and thus on the spectral shape in X-rays (see paper III, Sect. 2.1.2). Their impact however in the case of magnetically dominated (JED) disks remains to be investigated. Indeed, it remains to be checked whether or not these winds are being launched below $r_J$, i.e. within the JED portion of the disk. If this is actually confirmed, then a very interesting question arises: are those winds only the radiative signature of the base of the jets (low velocity and still massive outflow) or another, different, dynamical component making the transition from the outer standard accretion disk to the innermost jet-emitting disk? Although these observations do not involve \gx , this is a question that will require investigations in the future.

Last but not least, although timing properties are an important feature of X-ray binaries’ behavior, we did not discuss them here. Investigating hard/soft lags \citep[see][for a recent review]{2014A&ARv..22...72U} and quasi-periodic oscillations \citep[QPOs, see][]{Zhang13, Motta16} requires us to dig into the details of each observed spectrum. Such studies are out of the scope of the present paper. Our view is, however, consistent with multiple timing properties. First, the presence of an abrupt transition in disk density is a highly favorable place for the productions of QPOs within different mecanisms. Such a radius could be associated with the location of some specific instability \citep{Tagger99,Varniere12}, the transition from the outer optically thick to the inner optically thin accretion flow \citep[e.g.][]{Giannios04}, or the outer radius of the inner ejecting disk \citep[e.g.][]{Cabanac10}. Investigations will be presented in a forthcoming paper. Whether or not, however, such high density break is stable and produces QPOs is a question that needs to be addressed. We note for instance that high frequency QPOs are indeed observed in 3D GRMHD simulations, at the radial transition between the inner magnetically arrested disks (MAD) or magnetically choked accretion flows (MCAF) and the outer standard accretion disk \citep[see][and references therein]{McKinney12}. We also expect such a situation to arise in our case since the magnetic properties of MAD/MCAF are very close to that of the jet-emitting disk: a vertical magnetic field near equipartition, near-Keplerian accretion (beyond the plunging region), and a global accretion torque due to the presence of outflows. Second, while the production of time lags is usually associated with the presence of a corona/lampost geometry \citep[][Fig.~1]{2014A&ARv..22...72U}, very recent studies have shown that a radial stratification of the disk can be used to explain timing properties \citep{2018MNRAS.480.4040M, 2018arXiv181106911M}. This is very promising and fits surprisingly well with our own framework. These two timing questions will be raised and further studied in forthcoming works using dedicated observations.

\section{Conclusion}

Using the jet-emitting disk/standard accretion disk (JED--SAD) description for the inner regions of XrBs, we were able to replicate several observational diagnostics along the whole 2010--2011 cycle of \gx to a very good agreement. These diagnostics are the 3--200~keV total luminosity $L_{3-200}$, the power-law luminosity fraction $PLf = L_{pl}/L_{3-200}$, and the power-law index $\Gamma$. 
The observational X-ray constraints presented in this paper were derived from the \textit{RXTE}/PCA observations of \gx. The spectral parameters we use were provided by \citet{Clavel16} and directly compared to the theoretical ones. Those were obtained following the methods and limitations described in papers~II and III. The unknown dynamical parameters, allowed to vary along the cycle, are the JED--SAD transition radius $r_J$ and the disk accretion rate onto the black hole $\dot m_{in}$. 

Our present work followed a three steps process. We first looked for the best \pair light curves that could reproduce all X-ray diagnostics. We then showed, using a simple self-absorbed synchrotron emission model for the jet, that the model predicts a radio light curve qualitatively consistent with the \citet{Corbel13} observations of \gx at that same epoch. We realized that radio observations introduce a strong constraint on the JED--SAD transition radius $r_J$ during hard states, a constraint much more stringent than that provided by the X-ray data alone. 
As a last step, the radio luminosity has thus been included within the fits to constrain the disk dynamical state \pair accordingly with the X-ray spectral shape. This procedure greatly improved the agreement between the model and the radio observations, mostly by changing the magnitude and temporal behavior of $r_J$. Obviously, the parameters derived from X-ray observations can be strongly dependent on the spectral model selected and on the analysis procedures. We estimated the systematic errors introduced and in particular those due to the reflection component that is not yet included in our model. It turns out that they are consistent with the small discrepancies found between the model and the observations. We are thus confident on the robustness of our results, namely the reproductibility of the full XrB outburst and the trends derived for the model parameters \rj and \mdotin .

In order to fully understand the behavior of an outbursting XrB, one needs to explain both X-ray spectral changes and radio emission properties. A full cycle draws therefore a trajectory in a 3D space, made of the DFLD plus the radio flux axis. Ultimately, our approach can be thought as an effort to map this 3D evolutionary track into a 2D plot $r_J (\dot m_{in})$. If, indeed, a figure such as the one presented in Fig.~\ref{fig:ProcB_res}, bottom right panel, appears to be generic for a given object, then a dynamical explanation needs to be proposed. It will be a complex task since the evolution time scales involved are extremely long compared to any local timescale.

\begin{acknowledgements}
 We thank the anonymous referee and Joey Neilsen for their helpful comments and careful reading of the manuscript. The authors acknowledge funding support from the french research national agency (CHAOS project ANR-12-BS05-0009, http://www.chaos-project.fr), \textit{centre national d'études spatiales} (CNES), and the \textit{programme national des hautes énergies} (PNHE). This research has made use of data, software, and/or web tools obtained from the high energy astrophysics science archive research center (HEASARC), a service of the astrophysics science division at NASA/GSFC. Figures in this paper were produced using the \textsc{matplotlib} package \citep{plt}.
\end{acknowledgements}

\bibliographystyle{aa} 
\bibliography{ADSbibnew.bib}

\end{document}